\documentclass[aps,pra,twocolumn,bibnotes,notitlepage,longbibliography,superscriptaddress]{revtex4-1}
\usepackage[table,dvipsnames]{xcolor}
\usepackage{graphicx}
\usepackage{amsmath,amsfonts,bbm,bm,dsfont}
\usepackage{hyperref}
\hypersetup{colorlinks=true,linkcolor=blue,citecolor=magenta}
\usepackage[capitalise]{cleveref}

\usepackage{array}
\newcolumntype{L}{>{\centering\arraybackslash}m{3cm}}
\usepackage{makecell}

\newcommand\Tstrut{\rule{0pt}{2.9ex}}         
\newcommand\Bstrut{\rule[-1.2ex]{0pt}{0pt}}   

\usepackage[utf8]{inputenc}
\usepackage{newtxtext,newtxmath}
\usepackage[T1]{fontenc}
\usepackage{microtype}
\usepackage{xspace}
\xspaceaddexceptions{]\}}

\makeatletter
\g@addto@macro\bfseries{\boldmath}
\makeatother

\usepackage{parskip}

\newcommand{\Nmax}{\ensuremath{N_{\rm max}}}

\usepackage[tooltip=true]{acro}
\newcommand{\defacro}[3]{\DeclareAcronym{#1}{
  short = #2,
  long = #3
}}
\defacro{cov19}{COVID-$19$}{Coronavirus Disease $2019$}
\defacro{rtpcr}{RT-PCR}{Reverse Transcription -- Polymerase Chain Reaction}
\defacro{icmr}{ICMR}{Indian Council of Medical Research}
\defacro{lod}{LOD}{Limit of Detection}
\defacro{who}{WHO}{World Health Organization}

\date{\today}

\begin{document}

\title{Implementing Stepped Pooled Testing for Rapid COVID-19 Detection}

\author{Abhishek Srivastava}
\email{as2693@cornell.edu}
\affiliation{Minneapolis, MN 55414, USA}
\author{Anurag Mishra}
\affiliation{Los Angeles, CA 90048, USA}
\author{Trusha Jayant Parekh}
\affiliation{Mumbai, MH 400063, India}
\author{Sampreeti Jena}
\affiliation{Biochemistry, Molecular Biology, and Biophysics, University of Minnesota Twin
Cities, Minneapolis, MN 55414, USA}

\begin{abstract}
  \ac*{cov19}, a viral respiratory pandemic, has rapidly spread throughout the globe.
  Large scale and rapid testing of the population is required to contain the disease, but
  such testing is prohibitive in terms of resources, cost and time. Recently RT-PCR based
  pooled testing has emerged as a promising way to boost testing efficiency. We introduce
  a stepped pooled testing strategy, a probability driven approach which significantly
  reduces the number of tests required to identify infected individuals in a large
  population. Our comprehensive methodology incorporates the effect of false negative and
  positive rates to accurately determine not only the efficiency of pooling but also it's
  accuracy. Under various plausible scenarios, we show that this approach significantly
  reduces the cost of testing and also reduces the effective false positive rate of tests
  when compared to a strategy of testing every individual of a population. We also outline
  an optimization strategy to obtain the pool size that maximizes the efficiency of
  pooling given the diagnostic protocol parameters and local infection conditions. 
\end{abstract}

\maketitle
\setlength{\parskip}{2mm plus0.5mm minus0.5mm}

\onecolumngrid
\section{Introduction}
\label{sec:intro}

\ac{cov19}, a viral infectious respiratory illness, has recently emerged as a major threat
to public health and economic stability in countries around the world. It has spread globally
at an alarming pace and \ac{who} has declared it a pandemic. In absence
of a cure or a vaccine, large scale testing and quarantine is recognized as one of the
most effective strategies for containing its spread. While there are various known
diagnostic methods for \ac{cov19} including nucleic acid testing, protein testing and
computed tomography~\cite{udugama_2020_diagnosingcovid19}, they can be extremely
prohibitive in terms of cost and time. Pooled testing is a promising strategy to boost testing
efficiency. In pooled testing, several samples from each patient are divided and grouped
into various pools and the pool is then tested for the disease. If the pool tests
negative, each sample of the pool must be negative too. This basic idea reduces the
overall cost and time of testing large populations.

Pooled testing was first proposed during
World~War~II~\cite{dorfman_1943_detectiondefective} and has been a part of diagnostic
methodology ever since~\cite{bilder_2012_pooledtestingprocedures}. It has since been
employed several times to test for infections ranging including
Malaria~\cite{taylor_2010_highthroughputpooling}, Flu~\cite{arnold_2013_evaluationpooling}
and HIV~\cite{litvak_1994_screeningpresence,nguyen_2019_methodologyderiving}. One of the
first implementations of laboratory pooled testing for \ac{cov19} was demonstrated by Yelin
et al~\cite{yelin_2020_evaluationcovid19} for pools as large as 32 or 64 samples. Today,
Physicians and Public Health Officials from India~\cite{ht_2020_indiaasseses,
  goswami_2020_wewill,verma_2020_groupresearchers,perappadan_2020_icmrsuggests,
  presstrustofindia_2020_covid19pool,presstrustofindia_2020_centreallows,
  kaunainsheriffm_2020_24hrshifts} and many other countries around the
globe~\cite{technion_2020_poolingmethod, jeffay_2020_easeglobal,stone_2020_nebraskapublic,
  goetheuniversityfrankfurt_2020_coronapool,ghanaweb_2020_wepool} are using pooled testing
for determine the spread of this pandemic in a rapid and cost efficient manner.

A variety of different strategies have been proposed over the past several years to
implement pooled testing~\cite{dewolff_2020_evaluationpoolbased,
  theagarajan_2020_grouptesting}. They can be broadly classified into two types: adaptive
and non-adaptive. Adaptive methods~\cite{theagarajan_2020_grouptesting,
  shani-narkiss_2020_efficientpractical,noriega_2020_increasingtesting,
  bergel_2020_variablepool,zhu_2020_noisypooled,
  eberhardt_2020_multistagegroup,narayanan_2020_poolingrtpcr,
  ben-ami_2020_pooledrna,hanel_2020_boostingtestefficiency} employ a sequential testing
approach, thus requiring fewer number of total tests but more time as each step of testing
informs the next. On the other hand, non-adaptive pooled testing
methods~\cite{ben-ami_2020_pooledrna,taufer_2020_rapidlargescale,
  sinnott-armstrong_2020_evaluationgroup} usually involve a matrix type pooling that
allows for simultaneous testing of several pools whose results are then collated to
pinpoint to infected samples. These methods are faster but can require a greater number of
tests in total. While many of these methods might be mathematically efficient, their
practical implementation is usually
challenging~\cite{yelin_2020_evaluationcovid19,ben-ami_2020_pooledrna} and limits the
complexity that can be incorporated, no matter the benefits. Hence, it is imperative to
modify and verify any proposed method according to clinical constraints.

Here, we present a probability driven pooled testing approach that can significantly reduce
the number of tests required to identify infected patients in large populations. The
method divides and tests pools of samples in a hierarchical (stepped) manner. This
approach is general enough to not be limited to \ac{cov19} alone and can be applied to other
infectious scenarios with minor modifications. The mathematical model used for
implementing and optimizing this strategy is presented along with representative results
for various probable real-life scenarios. Under various plausible scenarios, this strategy
reduces the cost of testing between $30\%$ to $90\%$ compared to a strategy of individually
testing everyone in a population and cuts the false positive rate up to one-third of an
individual test.

It can be used to rapidly determine the efficiency boost that can be obtained by pooling a
desired number of samples together if we know the accuracy of testing method and the rate
of infection in the population being tested. It can also suggest optimal pool size that
should be used to minimize the number of tests needed per 1000 people.

\begin{table*}[t]
  \centering
  \begin{ruledtabular}
    \begin{tabular}[t]{ccp{5in}}
                               &\thead{Parameter Name} & \textbf{Description} \\
      \hline
                               & \Tstrut $M$           & Pool (group) size \\
                               & $\beta$               & Fraction of the Population infected \\
      \textbf{Inputs}          & $f_{+}$               & False positives rate for a test \\
                               & $f_{-}$               & False negatives rate for a test \\
      \Bstrut                  & \Nmax                 & Maximum number of tests possible per patient \\
      \hline
      \Tstrut \textbf{Outputs} & $K$                   & Efficiency Amplification Factor compared to testing
                                                         individually. (This is also the average effective number of people
                                                         that can be tested per test) \\
                               & $F_{-}^{\rm pool}$    & False Negative rate for the complete stepped pooled testing
                                                         (Different from $f_{-}$)
    \end{tabular}
  \end{ruledtabular}
  \caption{List of parameters used for Analysis and Results}
  \label{tab:1}
\end{table*}

\section{Methodology}
\label{sec:methods}

The stepped pooled testing strategy is applicable to any testing method that involves sample
collection such as the \ac{rtpcr}~\cite{udugama_2020_diagnosingcovid19} test which is being
widely used for testing \ac{cov19}. We begin by assuming that the sample(s) collected from
the patients are enough for \Nmax\ tests only (for instance if we are able to collect $3$
swabs per patient then $\Nmax=3$). This number will determines the number of steps of the
stepped pooled testing strategy.

Our strategy extends the $2$-step model described
in~\citet{hanel_2020_boostingtestefficiency}. The stepped pooled testing strategy goes as
follows:
\begin{enumerate}
\item We test a pool of $M$ samples.
\item If the outcome is negative (not infected) we can surmise that all the $M$ samples in
  the pool are infection free. 
\item If the pooled sample is tested positive (infected), we split the samples from these
  $M$ patients into two sub pools of size $M/2$ each and repeat steps 1 and 2. It should
  be noted that at every step of this process we need to use a fresh sample from the
  patient to make new sub-pools because the sample from the previous step is not
  reusable. 
\item This process is repeated $\Nmax-1$ times, after which we are left with a single
  sample of the patients in the sub-pools. If a sub-pool at this stage yields positive for
  infection, we individually test every patient in this sub-pool. 
\end{enumerate}

\begin{figure*}[t]
  \centering
  \includegraphics[scale=1.10]{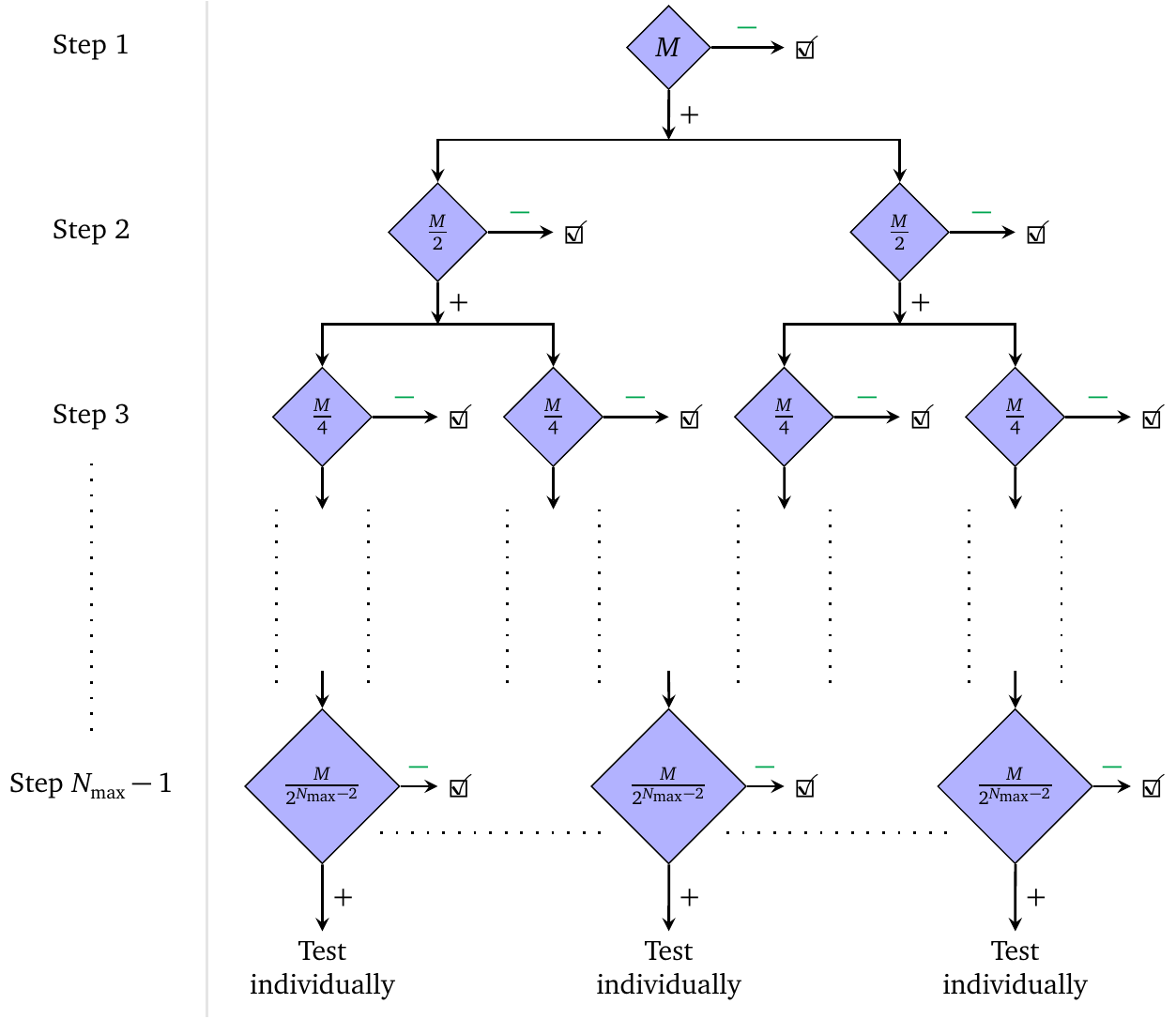}
  \caption{Flowchart (Decision Tree) representing the testing method for a pool of $M$
    samples. If a pool sample is tested negative $\textcolor{Green}{(-)}$, the procedure
    is stopped for that pool sample.}
  \label{fig:1}
\end{figure*}

It can be observed that this strategy is most effective when the the pool size $M$ is an \textit{integer multiple} of $M/2^{\Nmax-2}$. The initial size of the pool $M$ can be optimized to maximize the effective number of people tested per test or equivalently, minimize the number of tests needed per $1000$ people. A flowchart for this strategy is shown in~\cref{fig:1}.

Probabilistic calculations along this tree enable us to estimate the expected number of
tests to be done for a pool of given size as well as the overall chances of false
negatives.  The probability of a pool of $M$ samples being infected (i.e. at least $1$ out
of $M$ positive) is
\begin{equation}
  \label{eq:1}
  p(M) = 1 - \left( 1 - \beta \right)^{M} \ .
\end{equation}
The probability of the pooled testing positive is~\footnote{A pool that has infected samples may
  not necessarily \emph{test} as positive because the test has a non-zero false negative
  and false positive rates. Hence $G_{+}(M)$ is not the same as $p(M)$}
\begin{equation}
  \label{eq:2}
  G_{+}(M) = p(M)(1-f_{-}) + (1-p(M))f_{+} = p(M)(1-f_{-}-f_{+}) + f_{+} \ .
\end{equation}
Note that we have assumed that the false negative and positive rate for pool of samples is the same as that for a single sample. This can be justified based on the limits of detection for the commonly used RT-PCR protocols. Please refer to~\cref{app:LOD_FN} for details.

Following the flowchart in~\cref{fig:1}, we can deduce that $T(M)$, the expected number of
tests for a pool of size $M$, is given by a recursive function $\hat{Z}(M,s)$ that
terminates when we get to \Nmax\ steps:
\begin{align}
  \label{eq:3}
  T(M) =&\ \hat{Z}(M,\Nmax)    \\
  \hat{Z}(m,s)=&\
    \begin{cases}
      1 + 2G_{+}\hat{Z}(m/2,s-1) & \text{for}\ s>1 \\
      m & \text{for}\ s=1 \
    \end{cases}
\end{align}
Here $m$ denotes the subpool size and $s$ denotes the step number.

It follows that the number of persons per test, which we call the test efficiency
amplification $K$, is given by
\begin{equation}
  \label{eq:4}
  K = \frac{M}{T} \ .
\end{equation}

Correspondingly, the number of tests needed per 1000 people is
\begin{equation}
  \label{eq:6}
  T_{1000} = \frac{1000}{K} = \frac{1000T}{M} \ .
\end{equation}

The total probability of showing a false positive at the end of all steps can also be
calculated using a recursive formula. To better understand the calculation for this step,
it helps to write the probabilities at each step as shown in \cref{fig:5}. The recursive
formula for the pooled test false negative $F_{-}^{\rm pool}$ can then be written as
\begin{align}
  \label{eq:5}
  F_{-}^{\rm pool} =&\ \hat{U}(M,\Nmax) \\
  \hat{U}(m,s) =&\
     \begin{cases}
     p(m)f_{-} + 2p(m)(1-f_{-})\hat{U}(m/2,s-1) & \text{for}\
     s > 1 \\
     f_{-}/2 & \text{for}\ s = 1 \ .
                  \end{cases}
\end{align}
Here $m$ denotes the subpool size and $s$ denotes the step number.

\section{Results}
\label{sec:results}

\Ac{icmr} recently published guidelines~\cite{prakash_2020_advisoryfeasibility} for pool
testing and suggested limiting the pool size $M$ to 5 to avoid dilution. \Ac{icmr} also
suggested a staggered approach to use of pooled testing: (a) for areas with infection rate
in the population less than $2\%$ pooled testing should be used, (b) For infection rate
between $2-5\%$, pooled testing should be used for community and asymptomatic patient
testing, and (c) for areas with infection rate > $5\%$, pooled testing should not be
used. We will use these numbers as a guide for demonstrating our method. It should be
noted that higher pool sizes, up-to $M=64$, have been reported in other
studies~\cite{yelin_2020_evaluationcovid19}. These are also in agreement with our calculations regarding limits of detection (See~\cref{app:LOD_FN}).

\begin{figure*}
  \centering
  \includegraphics[width=\textwidth]{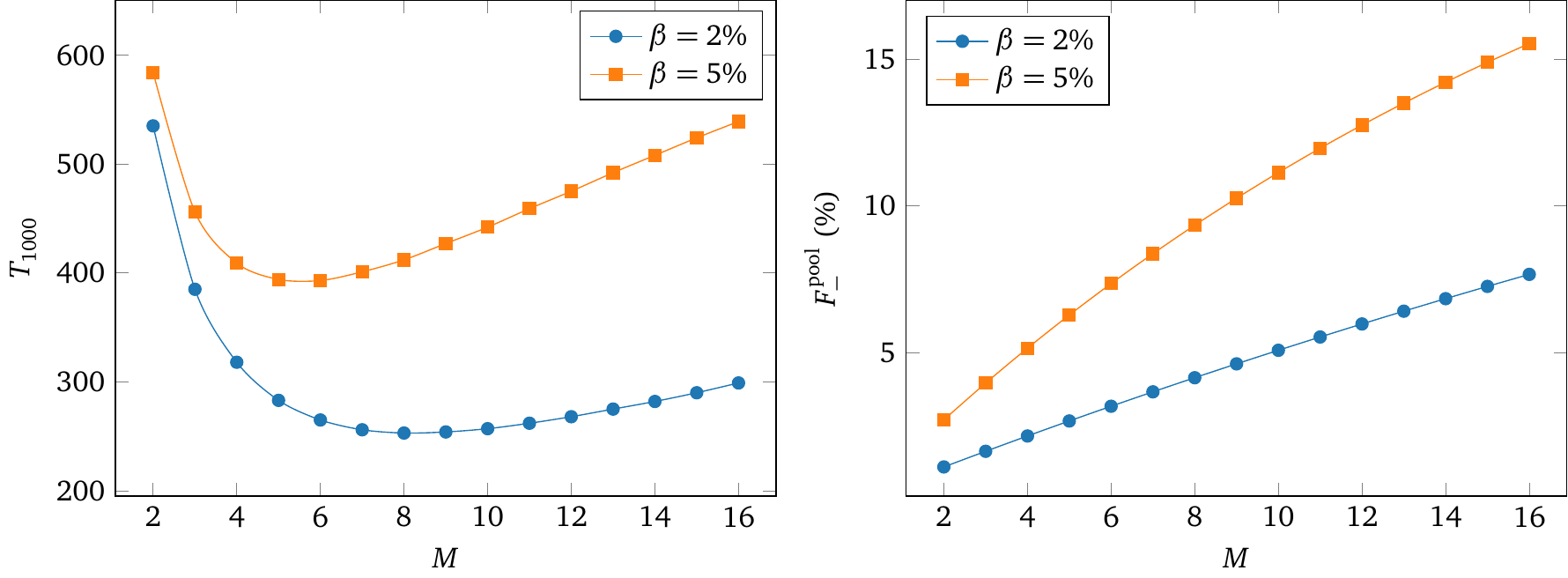}%
  \caption{\textbf{Effect of population infection rate $\beta$.} Tests required per 1000
    people (left) and pool test false negative percentage (right) as a function of pool
    size. We assume number of steps $\protect\Nmax=2$, a false positive rate of
    $f_+=0.12\protect\%$ and a false negative rate of $f_-=15\protect\%$.}%
  \label{fig:2}%
\end{figure*}

\begin{figure*}
  \centering
  \includegraphics[width=\textwidth]{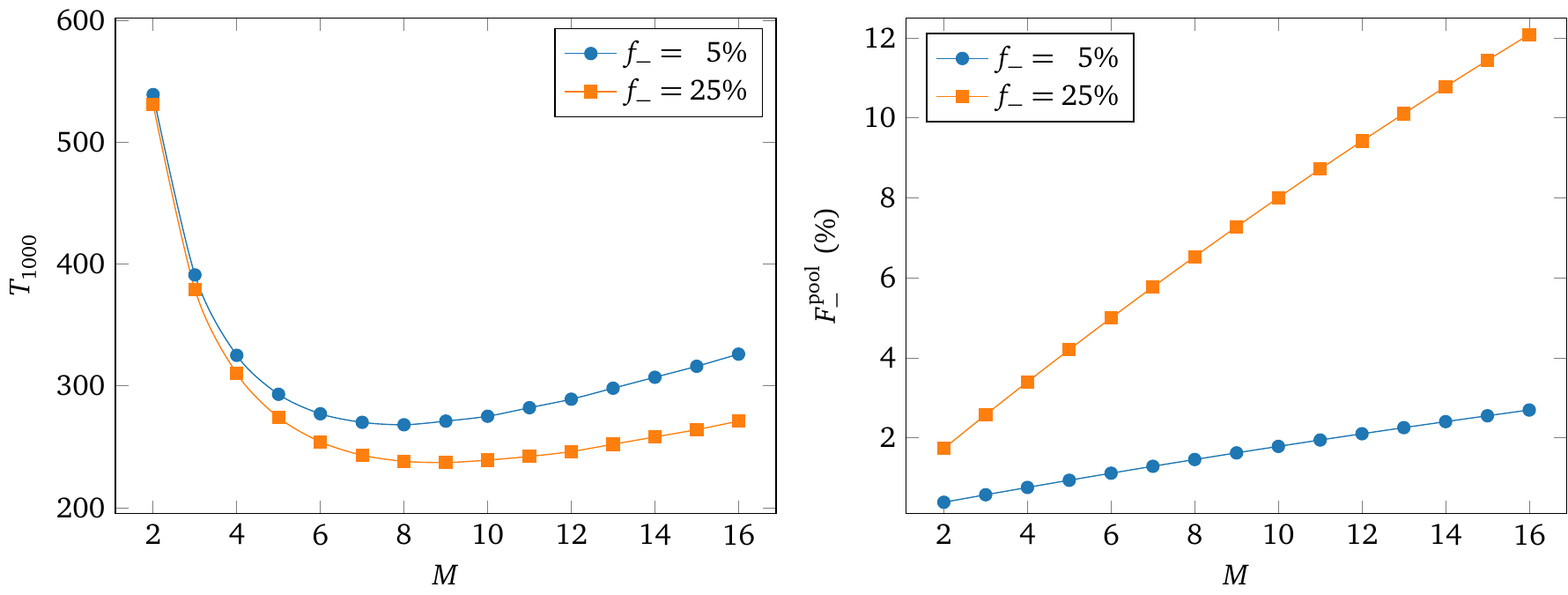}%
  \caption{\textbf{Effect of false negative rate $f_{-}$.} Tests required per 1000 people
    (left) and pool test false negative percentage (right) as a function of pool size. We
    assume number of steps $\protect\Nmax=2$, an infection rate $\beta=2\protect\%$ and a
    false negative rate $f_-=15\protect\%$.}%
  \label{fig:3}%
\end{figure*}

\begin{figure*}
  \centering \includegraphics[width=\textwidth]{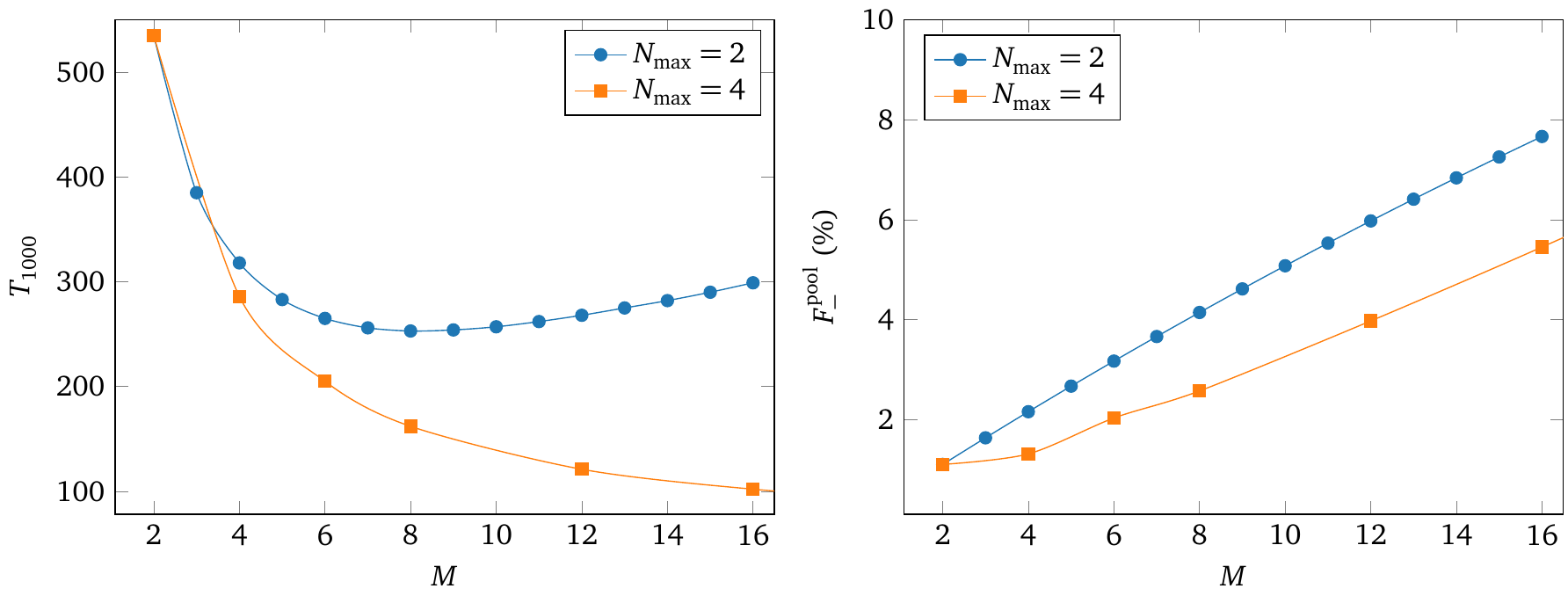}%
  \caption{\textbf{Effect of number of steps $\Nmax$.} Tests required per 1000 people
    (left) and pool test false negative percentage (right) as a function of pool size for
    $\Nmax=2$ and $\Nmax=4$. We assume an infection rate of $\beta=2\protect\%$, a false
    negative rate $f_-=15\protect\%$ and a false positive rate $f_{+}=0.12\protect\%$.}
  \label{fig:4}%
\end{figure*}

In~\cref{fig:2,fig:3,fig:4}, we show the results for a representative set of
parameters. We find that the number of tests per $1000$ people decreases and the false
negatives increases as we make the pool size larger. However, there is an optimum pool
size that achieves maximum efficiency (i.e. minimum $T_{1000}$).

\Cref{fig:2} reveals that for the same pool size, a higher infection rate population
requires more tests and will have an overall lower accuracy (higher false negative
rate). This is consistent with what we would expect clinically. In~\cref{fig:3}, we obtain
the effect of false negative rate on stepped pooled testing. Interestingly, a diagnostic
test with higher false negative would go through more samples in a fewer number of tests
but at the cost of overall higher pool test false negative making this trade-off possibly
undesirable. 

In~\cref{fig:4}, we see the effect of the number of steps, $\Nmax$ (also the number of
samples per patient) on the pooling strategy. Similar to the previous two parameter
sweeps, we notice that the test required per $1000$ people shows a non-monotonic behavior
and has an optimal pool size for which the pooling is most efficient (Note that for
$\Nmax=4$, $T_{1000}$ minimizes at $M=48$ which is beyond the visible horizontal axis). On
the other hand, the false negative rate steadily increases but still remains below the
false negative rate of a single test. It is obvious that using multiple samples
significantly reduces the number of tests needed without compromising the overall false
negative of the pooling strategy.

\Cref{tab:2} summarizes the results for a broad set of plausible scenarios to demonstrate
the efficiency of this strategy. In addition to predicting the efficiency and accuracy of
different pooling strategies, we can also this method to calculate the optimal pool size
that leads to the least number of tests (i.e. minimizes
$T_{1000}$). \Cref{fig:2,fig:3,fig:4} clearly demonstrate the existence of such an
optimum. In~\cref{tab:3}, we show various possible testing scenarios and the corresponding
optimal pool size. The results in this section show that stepped pooled testing can reduce
the overall pool false negative rate below the false negative rate of an individual test.

\begin{table*}[h]
  \centering
  \begin{ruledtabular}
  \begin{tabular}{lccccc}
    \thead{Infection Rate ($\beta$)} & \thead{Number of Steps \Nmax} & \thead{Pool Size ($M$)} & \thead{Test Needed Per\\ 1000 People $T_{1000}$} & \thead{Pool Test False\\ Negative $F_{-}^{\rm pool}$ \%} & \thead{Testing Cost\\ Reduction (\%)} \\
    \hline
    \Tstrut      &  2 & 2 & 535 & 1.10  & 46.5 \\
    \            &  2 & 3 & 385 & 1.63  & 61.5 \\
    2\% (Low)    &  2 & 5 & 283 & 2.67  & 71.7 \\
    \            &  3 & 4 & 286 & 1.31  & 71.4 \\
    \            &  3 & 6 & 205 & 2.03  & 79.5 \\
    \Bstrut      &  4 & 8 & 162 & 2.57  & 83.8 \\
    \hline
    \Tstrut      & 2  & 2 & 584 & 2.70  & 41.6 \\
    \            & 2  & 3 & 456 & 3.96  & 54.4 \\
    5\% (Medium) & 2  & 5 & 394 & 6.27  & 60.6 \\
    \            & 3  & 4 & 343 & 3.64  & 65.7 \\
    \            & 3  & 6 & 267 & 5.76  & 73.3 \\
    \Bstrut      & 4  & 8 & 224 & 7.13  & 77.6 \\
    \hline
    \Tstrut      & 2  & 2 & 663 & 5.27  & 33.7 \\
    \            & 2  & 3 & 565 & 7.52  & 43.5 \\
    10\% (High)  & 2  & 5 & 549 & 11.36 & 45.1 \\
    \            & 3  & 4 & 445 & 8.24  & 55.5 \\
    \            & 3  & 6 & 392 & 13.02 & 60.8 \\
    \            & 4  & 8 & 341 & 16.52 & 65.9 \\
  \end{tabular}
\end{ruledtabular}
\caption{\textbf{Testing cost reduction from stepped pool strategy.} We show the overall
  testing cost reduction for various plausible scenarios outlined by \protect\ac{icmr}. We
  assume a false negative rate $f_{-}=15\protect\%$ and a false positive rate
  $f_{+}=0.12\protect\%$.}
\label{tab:2}
\end{table*}

\begin{table}
  \centering
  \begin{ruledtabular}
  \begin{tabular}{lr|cccc}
    \ & \ & \thead{Optimal Pool\\ Size ($M$)} & \thead{Test Needed Per\\ 1000 People $T_{1000}$} & \thead{Pool Test False\\ Negative $F_{-}^{\rm pool}$ \%} & \thead{Testing Cost\\ Reduction (\%)} \\
    \hline
    \Tstrut                       &  2\,\%\;\; & 8  & 253 & 4.14  & 74.7 \\
    Infection Rate ($\beta$)      &  5\,\%\;\; & 6  & 393 & 7.35  & 60.7 \\
    \Bstrut                       & 10\,\%\;\; & 4  & 544 & 9.54  & 45.6 \\
    \hline\hline
    \Tstrut                       &  5\,\%\;\; & 8  & 268 & 1.46  & 73.2 \\
    False negative                & 15\,\%\;\; & 8  & 253 & 4.14  & 74.7 \\
    rate ($f_{-}$)                & 30\,\%\;\; & 9  & 229 & 8.48  & 77.1 \\
    \Bstrut                       & 40\,\%\;\; & 10 & 211 & 11.71 & 78.9 \\
    \hline\hline
    \Tstrut                       &      2\;\; & 8  & 253 &  4.14 & 74.7 \\
    Number of steps (\Nmax)       &      3\;\; & 18 & 122 &  6.96 & 87.8 \\
    \                             &      4\;\; & 32 &  81 & 12.07 & 91.9
  \end{tabular}
\end{ruledtabular}
\caption{\textbf{Optimal pool size $M$ under various scenarios.} Unless specified in the
  first column, we use number of steps $\Nmax=2$, an infection rate of $\beta=2\protect\%$, a false
  negative rate $f_-=15\protect\%$ and a false positive rate $f_+=0.12\protect\%$.}
\label{tab:3}
\end{table}

\section{Conclusion}
\label{sec:conclusion}

We propose a new stepped pooled testing strategy that can significantly reduce the cost of
testing a large population. The strategy also reduces the chances of false negative in
almost all scenarios because an infected patient's sample is likely to be tested multiple
times. Even in the simplest case with two samples per individual (i.e. two steps, also called \emph{Dorfman} Pooling~\cite{dorfman_1943_detectiondefective}) and an initial pool size of $2$, we can significantly reduce the number of tests required per 1000 individuals, by up to $33.7\%$ for populations with a high infection rate and up to $46.5\%$ for populations with a low infection rate. As the number of steps and initial pool size is increased, the testing efficiency progressively improves, albeit at the cost of slightly higher false negative rate. Never the less, barring the cases with very high infection rate, the pooled false negative rate is still below that of an individual test.

Based on our results, we make several suggestions about the effective pool size and the
number of samples that should be collected from an individual. This methodology should be
customized dynamically and regularly based on evolving local levels of infection. Most
significant benefits of this strategy can be realized by collecting $2$ or $3$ samples
from each individual and pooling them into groups of $4$ to $6$. Increasing the number of
steps \Nmax\ means collecting more samples from each patient being tested. Hence, the
value of \Nmax\ should be chosen pragmatically based on consultation with the physician or
health professional. Finally, we note that machine learning methods may be implemented to
utilize data collected on disease spread and dynamically adapt this strategy for maximum
efficiency. We leave this as a topic for future research.

\acknowledgments
The authors are thankful to Dr. Saumya Srivastava, MBBS and Vertika Srivastava for useful discussions, and to Dr. Hanel for providing more details about his model via email.
\newcommand{\Eprint}[2]{}
\twocolumngrid

\onecolumngrid
\appendix

\section{Diagnostic limit considerations for pooled testing with RT-PCR}
\label{app:LOD_FN}

One of the key advantages of real-time PCR assays utilizing target sequence specific
primers (as is the case with all \ac{cov19} test kits) is their wide dynamic range. This
enables the analysis of samples with widely varying levels of target RNA. The resolving
power of \ac{rtpcr} is mostly limited by the efficiency of RNA-to-cDNA conversion, a real
concern when the target RNA is scarce. Thus, determination of the \ac{lod}-- by performing
serial dilutions of the positive control sample and obtaining standard curves-- is a
critical step in the validation of any testing kit/protocol. The highest dilution of the
standard curve, provided in the assay performance evaluation report of any \ac{rtpcr} assay
kit, delineates the lowest concentration that can be quantified with confidence. Thus,
pooling patient samples as proposed by the current model is unlikely to influence the
probability of a false negative prediction by the assay if the effective target
concentration is maintained above the \ac{lod}. However, if the intensity values recorded are
comparable to that of the \ac{lod}, they should be recorded only as a qualitative (yes/no)
prediction~\cite{bustin2004_pitfalls}.

Target RNA selection plays a big role in the assay sensitivity. These include
RNA-dependent RNA polymerase (RdRp), hemagglutinin-esterase (HE), and open reading frames
ORF1a and ORF1b. \Acf{who} recommends a first line screening with the E gene assay followed
by a confirmatory assay using the RdR p gene. Tang \emph{et
  al.}~\cite{tang2020_labdiagnosis} developed and compared the performance of three novel
real-time RT-PCR assays targeting the RdRp/Hel, S, and N genes of SARS-CoV-2. Among them,
the COVID-19-RdRp/Hel assay had the lowest limit of detection in vitro and higher
sensitivity and specificity.

In this section, we will calculate the maximum possible pool size ($M^{*}$) that is
consistent with the \ac{lod} of current \ac{cov19} tests.

\begin{table*}
  \centering
  \begin{ruledtabular}
  \begin{tabular}{llc}
    \ \hspace{2cm}   & \textbf{Specimen Type}\hspace{3cm} & \thead{Mean (range) viral load (RNA copies/mL) in RdRp-P2-negative\\but COVID-19-RdRp/Hel-positive specimens}\hspace{2cm} \\
    \hline
    \ \hspace{2cm}   & Respiratory tract                  & $4.33\times10^4(2.85\times10^3-4.71\times10^5)$ \\
                     & \hspace{3mm} NPA/NPS/TS            & $1.74\times10^4(2.85\times10^3-8.40\times10^4)$ \\
                     & \hspace{3mm} Saliva                & $5.32\times10^4(1.74\times10^3-4.71\times10^5)$ \\
                     & \hspace{3mm} Sputum                & N/A                                             \\         
                     &                                    &                                                 \\            
                     & Non-respiratory tract              & $7.06\times10^3(2.21\times10^2-1.67\times10^4)$ \\
                     & \hspace{3mm} Plasma                & $7.86\times10^3(2.21\times10^2-1.67\times10^4)$ \\
                     & \hspace{3mm} Urine                 & N/A                                             \\         
                     & \hspace{3mm} Feces/rectal swabs    & $4.38\times10^3(1.54\times10^3-6.69\times10^3)$ \\
                     &                                    &                                                 \\
                     & Total                              & $3.21\times10^4(2.21\times10^2-4.71\times10^5)$ \\
  \end{tabular}
\end{ruledtabular}
\caption{Viral Load in respiratory and non-respiratory specimens. Reproduced from Chan \emph{et al.}~\cite{chan2020_improveddiagnosis}}
\label{tab:4}
\end{table*}

\noindent\textbf{Calculation}.-- The \ac{lod} of the COVID-19-RdRp/Hel assay is $11.2$ RNA
copies/reaction~\cite{tang2020_labdiagnosis}. Assuming a reaction volume of $25\ \mu$L, this
is equivalent $448$ RNA copies in one mL sample.  From~\cref{tab:4}, we find that the mean
viral load for nasopharyngeal/nasal swabs is $1.74\times10^4$ RNA copies/mL.  Assuming a
pool size of $M$ samples with only one infected sample, and that samples are pooled
first followed by RNA extraction, the net effective viral load in the pooled sample will be
$(1.74/M)\times10^4$ copies/mL.  In the standardized protocol for RNA extraction and
\ac{rtpcr} procedure, $200$ $\mu$L of pooled sample is diluted with $250$ $\mu$L of solvent
and loaded for RNA extraction. Purified RNA is diluted into $50$ $\mu$L of solvent. $10$ $\mu$L
of diluted solution is used per well of PCR assay with a total reaction volume of $25$
$\mu$L~\cite{carter2020_assaytechniques}.\\

Thus, the net effective viral load per PCR well (in units of RNA copies/mL of solvent) is
\begin{equation}
  \label{eq:7}
  \frac{1.74\times10^4}{M}\times\frac{200}{450}\times\frac{450}{50}\times\frac{10}{25} \approx \frac{2.8\times10^4}{M} \ .
\end{equation}
This quantity should be greater than the minimum \ac{lod} of the test, which is 448 RNA
copies per one mL of solvent. Thus,
\begin{align}
  \label{eq:8}
  \frac{2.8\times10^4}{M} &\geq 448 \\
  \text{or,}\qquad\qquad M &\leq 62.5
\end{align}
Thus, the largest possible pool size consistent with \ac{lod} of a \ac{rtpcr} test is
$M^{*}=62$. This value is consistent with earlier
literature~\cite{yelin_2020_evaluationcovid19,narayanan_2020_poolingrtpcr}. 

\section{Pool Test False Negative}
\label{app:poolFN}
\begin{figure}[h]
  \centering
  \includegraphics[scale=1.10]{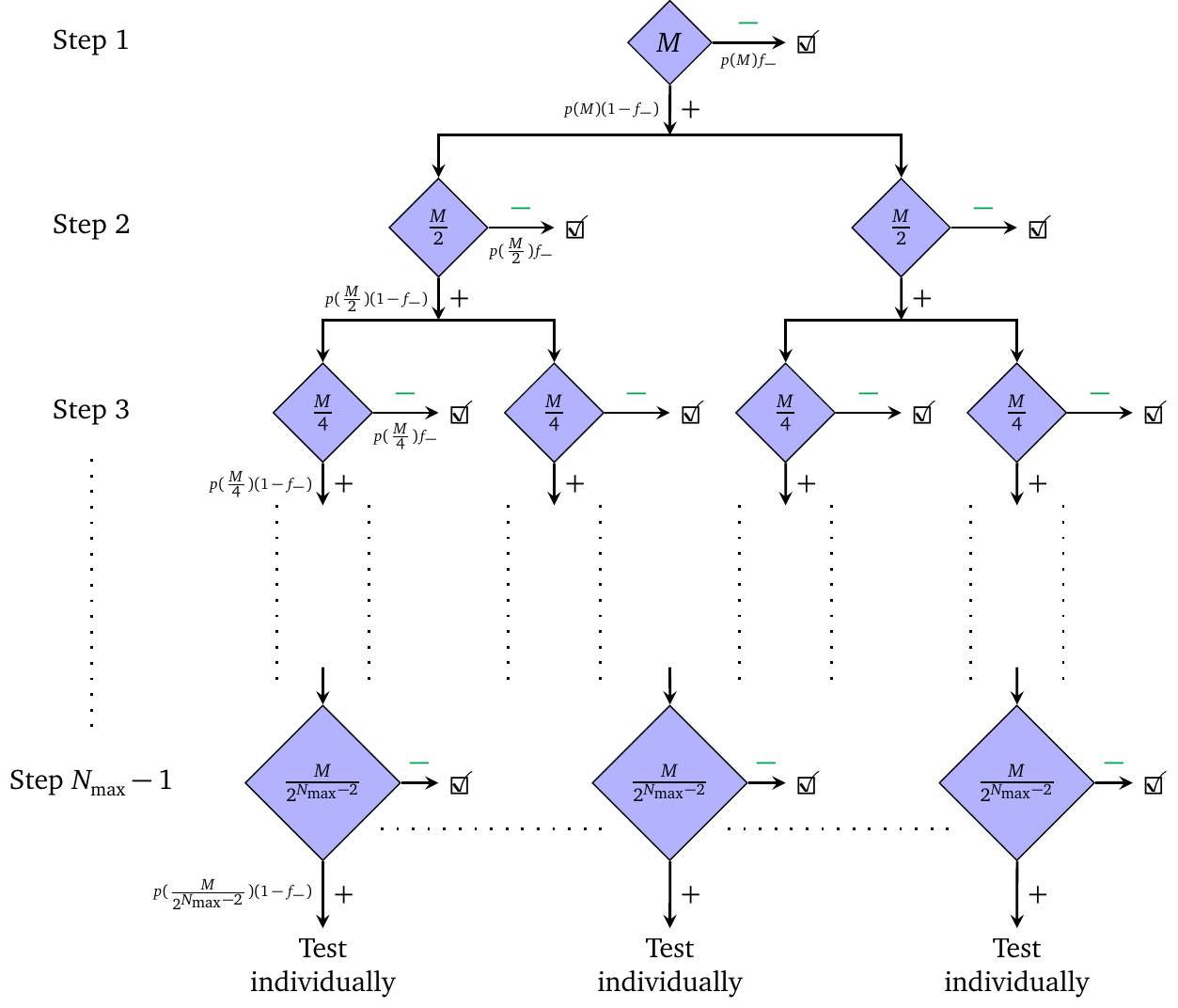}
  \caption{Flowchart showing the calculation for pool test false negative. The
    representative values are shown on the left side of the tree.}
  \label{fig:5}
\end{figure}

\end{document}